\documentclass{article}
\usepackage{spconf,amsmath,graphicx,hyperref}
\usepackage{cite}
\usepackage{amsmath,amssymb,amsfonts}
\usepackage{algorithmic}
\usepackage{graphicx}
\usepackage{textcomp}
\usepackage{xcolor}
\usepackage{hyperref}
\usepackage{subcaption}
\usepackage{url}
\usepackage{multicol}
\usepackage{multirow}
\usepackage{mdwlist}
\usepackage{enumitem}
\usepackage{amsthm}
\usepackage{booktabs}
\usepackage{algorithm}
\usepackage[switch]{lineno}
\usepackage{tabularx}
\usepackage[misc]{ifsym}

\newcommand{\Tref}[1]{\textcolor{black}{Table~\ref{#1}}}

\newcommand{\Fref}[1]{\textcolor{black}{Fig.~\ref{#1}}}

\def\BibTeX{{\rm B\kern-.05em{\sc i\kern-.025em b}\kern-.08em
    T\kern-.1667em\lower.7ex\hbox{E}\kern-.125emX}}

\makeatletter
\renewcommand{\maketag@@@}[1]{\hbox{\m@th\normalsize\normalfont#1}}%
\makeatother
\newcommand{\para}[1]{{\vspace{1.5pt} \bf \noindent #1 \hspace{2pt}}} 

\begin{document}
\title{DisSR: Disentangling Speech Representation for Degradation-Prior Guided Cross-Domain Speech Restoration}

\name{
    Ziqi Liang$^{1\dagger}$, Zhijun Jia$^{1\dagger}$, Chang Liu$^{2}$, Minghui Yang$^{1}$, Zhihong Lu$^{1}$, Jian Wang$^{1\ast}$\thanks{$\dagger$ Equal Contribution}\thanks{$^\ast$Corresponding author: Jian Wang, bobblair.wj@antgroup.com.}
    \vspace{-0.3cm} 
}
\address{
    $^{1}$ AntGroup, HangZhou, China \\
    $^{2}$ University of Science and Technology of China, Hefei, China }
    
\maketitle

\begin{abstract}
Previous speech restoration (SR) primarily focuses on single-task speech restoration (SSR), which cannot address general speech restoration problems. 
Training specific SSR models for different distortions is time-consuming and lacks generality.
In addition, most studies ignore the problem of model generalization across unseen domains. To overcome those limitations, we propose \textbf{DisSR}, a \underline{Dis}entangling \underline{S}peech \underline{R}epresentation based general speech restoration model with two properties: 1) Degradation-prior guidance, which extracts speaker-invariant degradation representation to guide the diffusion-based speech restoration model. 2) Domain adaptation, where we design cross-domain alignment training to enhance the model's adaptability and generalization on cross-domain data, respectively.
Experimental results demonstrate that our method can produce high-quality restored speech under various distortion conditions.
Audio samples can be found at \href{https://itspsp.github.io/DisSR}{https://itspsp.github.io/DisSR}.
\end{abstract}

\begin{keywords}
speech restoration, degradation-prior guidance, doamin adaption
\end{keywords}

\section{Introduction}
Speech signals collected from real-world scenarios often suffer from various distortions, including bandwidth limitations~\cite{HIFI}, background noise~\cite{UDiffSE}, reverberation~\cite{Dereverberation}, overdrive~\cite{selfremaster}, and other types of distortion~\cite{declipping}.
These factors can significantly reduce the perceptual quality of the input signal, making it challenging to understand the target speech.
In order to expand the application scope of speech processing, it is necessary to utilize low-quality speech, while speech restoration plays an important role.

Previous works in speech restoration mainly focus on dealing with only one type of distortion at a time. 
However, in the real world, speech signals can be degraded by several different distortions simultaneously, which means previous SSR systems oversimplify the speech distortion types \cite{BandwidthExtension1,audio_sr2}. To handle various degradations, several methods have been proposed to investigate this problem. VoiceFixer~\cite{voicefixer} proposes a generative model to eliminate multiple distortions simultaneously. 
It is trained using artificial paired data created by various distortions, but when faced with new distortion types, the model will forget prior knowledge, reducing the model's adaptability to certain restoration tasks.
Although \cite{selfremaster} presents a self-supervised speech restoration model without paired data, its restoration effect will be damaged when faced with out-of-domain low-quality speech.
\cite{cvpr_invar} have proven that domain-invariant features as a prior can improve the robustness of image restoration models. 
\begin{figure}[t]
    \centering
    \includegraphics[width=8.5cm]{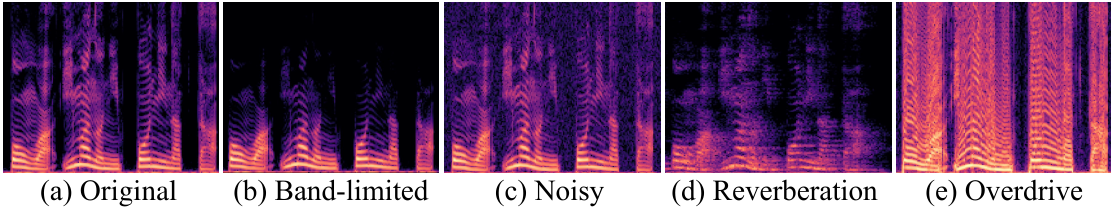}
    \vspace{-0.2cm}
    \caption{Different Distortions in Speech Signals} 
    \label{pretrain}
    \vspace{-0.5cm}
\end{figure}
Based on the above observations, it would be better to achieve degradation-prior guidance and domain adaptation.
The domain adaptation in speech restoration can be regarded as domain generalization for unseen speaker styles. We can extract speaker-invariant features as domain-invariant features to enhance the model generalization towards unseen speaker styles. 

In this paper, we propose a general speech restoration model with degradation-prior guidance to restore multiple speech distortions. Unlike the other methods \cite{BandwidthExtension1,audio_sr2,voicefixer} that explicitly estimate the degradation, our model learns a latent degradation feature which is speaker-invariant to distinguish distortions in the representation space rather than explicit estimation.  
Our contributions are as follows:

\begin{figure*}[h]
    \centering
    \includegraphics[width=17.0cm]{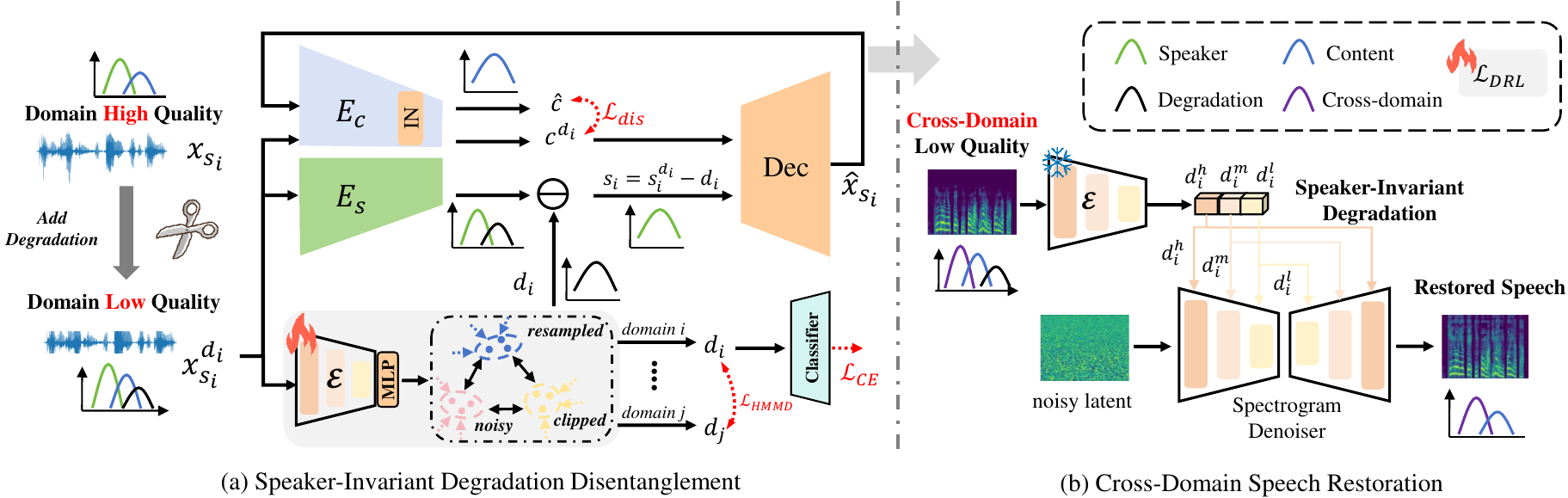}
    \caption{ Overall pipeline of DisSR: $c^{d_{i}}$ and $s_{i}^{d_{i}}$ are the content and speaker style extracted from input $x_{s_{i}}^{d_{i}}$. Instance Normalization (IN) can eliminate the global speaker style from $x_{s_{i}}^{d_{i}}$. $d_{i}$ is disentangled as degradation-prior to guide restoration. $c^{d_{i}}$ and $\hat{c}$ are the content from $x_{s_{i}}^{d_{i}}$ and predicted speech $\hat{x_{s_{i}}}$, respectively. $s_{i}$ is clean speaker style for reconstruction training.
    } 
    \vspace{-0.5cm}
    \label{stage1}
\end{figure*}

\begin{enumerate}
\item The hypothesis that distorted information exists in speaker style is proposed and experimentally verified. We obtain speaker-invariant degradation from speaker style through speech representation disentanglement.
\item We leverage the speaker-invariant degradation as a conditional prompt to guide the diffusion model, enabling it to achieve general speech restoration for various distortions.
\item We adopt cross-domain alignment training for improving model generalization on cross-domain data. Experiments demonstrate the superiority of general speech restoration.
\end{enumerate}

\section{DisSR}
\subsection{Background and hypothesis}

Previous work has proven that speech can be disentangled into speaker style features and content features \cite{one-shotvc,eadvc,Zhang,Liang0LQZW25}. 
We extract degradation representation from distorted speech's speaker style features as a domain-invariant features to guide speech restoration.
The correlation of degradation $d_{i}$ with speaker style $s_{i}$ and content $c$ is expressed as follow:
\begin{equation}
    \begin{split}
        x_{s}^{d_{i}} = x_{s}\otimes d_{i}=c \oplus s_{i}^{d_{i}}=c \oplus s_{i}\otimes d_{i}
    \end{split}
\end{equation}
where $x_{s}^{d_{i}}$ and $x_{s}$ represent distorted speech and high-quality speech respectively. $x_{s}^{d_{i}}$ is a coupling of $x_{s}$ and $d_{i}$, which can also be disentangled into content embedding $c$ and distorted speaker style embedding $s_{i}^{d_{i}}$, and further disentangle $s_{i}^{d_{i}}$ into clean speaker style embedding $s_{i}$ and $d_{i}$. $\oplus$ means content-style coupling, $\otimes$ means  style coupling.
We proved our hypothesis in Section 3.2 that the degradation in speech is mainly contained in speaker style, which is independent of content. 

\subsection{Speaker-invariant degradation disentanglement}
As shown in \Fref{stage1} (a), we use the pretrained content encoder $E_{c}$ and speaker style encoder $E_{s}$ to disentangle content $c^{d_{i}}$ and speaker style $s_{i}^{d_{i}}$ respectively \cite{adain-vc}.
For cross-domain speech restoration task, we design a degradation encoder $E_{\varepsilon}$ to extract speaker-invariant features $d_{i}$ as degradation-prior to guide diffusion-based restoration model. 
Each input low-quality speech $x_{s_{i}}^{d_{i}}$ passes through $E_{s}$ to obtain the speaker style $s_{i}^{d_{i}}$ mixed with degradation $d_{i}$, and then we can get clean speaker style with speaker-invariant features $d_{i}$:
\begin{equation}
    \begin{split}
        s_{i} = s_{i}^{d_{i}}-d_{i}= E_{s}(x_{s_{i}}^{d_{i}})-E_{\varepsilon}(x_{s_{i}}^{d_{i}})
    \end{split}
\end{equation}
where $i$ is the style identity, representing the domain of different data distributions. Then, $s_{i}$ and $c^{d_{i}}$ are fed to decoder to get the predicted speech $\hat{x}_{s_{i}}$.
Since $x_{s_{i}}^{d_{i}}$ and $\hat{x}_{s_{i}}$ have the same content information, their content embeddings are expect to be as closer as possible, which is followed as:

\begin{small}
\vspace{-0.2cm}
\begin{equation}
    \begin{split}
        \mathop{\min}_{E_{\varepsilon}(\cdot),E_{mlp}(\cdot)}\mathcal{L}_{\mathrm{dis}} &= \mathbb{E} \left[ \parallel E_{c}(\hat{x}_{s_{i}})-E_{c}({x}_{s_{i}}^{d_{i}}) \parallel_{1}^{1} \right] \\
    \end{split}
\end{equation}
\end{small}

\para{Degradation representation learning (DRL).} 
We design the DRL to disentangle degradation from speaker style. 
Let $d_{i}$ and $d_{j}$ represent distinct degradation types, while $s_{i}$ and $s_{j}$ represent different domains. Given a distorted speech $x_{s_{i}}^{d_{i}}$ as the query, other speech extracted from the same mini-batch can be considered as positive samples $x_{s_{j}}^{d_{i}}$. In contrast, distorted speech from other distortion types can be referred to as negative samples $x_{s_{i}}^{d_{j}}$. Then, we encode the query, positive and negative samples into degradation representations using a UNet-based degradation encoder. Similar to contrastive learning, $x_{s_{i}}^{d_{i}}$ is encouraged to be similar to $x_{s_{j}}^{d_{i}}$ while being dissimilar to $x_{s_{i}}^{d_{j}}$, which is defined as follows:

\begin{footnotesize}
\begin{equation}
    \begin{split}    
    \mathcal{L}_{\mathrm{DRL}}=-\log\frac{ \mathrm{exp}(\mathrm{sim}(x_{s_{i}}^{d_{i}} \cdot x_{s_{j}}^{d_{i}}))}{ \mathrm{exp}(\mathrm{sim}(x_{s_{i}}^{d_{i}} \cdot x_{s_{j}}^{d_{i} })) +\sum_{j=1}^{N}exp(sim(x_{s_{i}}^{d_{i}} \cdot x_{s_{i}}^{d_{j}})) }  
    \end{split}
\end{equation}
\end{footnotesize}


\begin{table*}[t!]
\caption{Evaluation results on cross-domain unseen speaker style, which includes all 6 studied distortions.}
\vspace{-3pt}
\label{audioSR}
\centering
\scalebox{0.82}{
\begin{tabular}{@{}c|ccc|ccc|ccc@{}}
\toprule
\cmidrule(r){1-10}
\multirow{2.5}{*}{  \textbf{Methods}} & \multicolumn{3}{c|}{\bfseries LibriTTS$\rightarrow$VCTK (EN)} & \multicolumn{3}{c|}{\bfseries LibriTTS$\rightarrow$AISHELL-3 (ZH)} & \multicolumn{3}{c}{\bfseries LibriTTS$\rightarrow$JSUT (JP)}\\ 
\cmidrule(l){2-10}  
&DNSMOS $\uparrow$ &PESQ-wb $\uparrow$ &MCD $\downarrow$ 
&DNSMOS $\uparrow$ &PESQ-wb $\uparrow$ &MCD $\downarrow$
&DNSMOS $\uparrow$ &PESQ-wb $\uparrow$ &MCD $\downarrow$  \\ 
\midrule
Unprocessed
& 2.76$\pm$0.13  & 1.94$\pm$0.13 & 14.20$\pm$0.12   
& 2.58$\pm$0.15  & 1.86$\pm$0.08  & 11.71$\pm$0.08  
& 3.15$\pm$0.09  & 2.12$\pm$0.12  & 12.63$\pm$0.09  \\
VoiceFixer \cite{voicefixer} 
& 3.45$\pm$0.12  & 2.37$\pm$0.11 & 8.97$\pm$0.08   
& 3.18$\pm$0.15  & 2.26$\pm$0.10  & 7.71$\pm$0.09  
& 3.15$\pm$0.09  & 2.12$\pm$0.12  & 8.20$\pm$0.13  \\
SelfRemaster \cite{selfremaster}  
& 3.52$\pm$0.16  & 2.49$\pm$0.08 & 8.45$\pm$0.11  
& 3.30$\pm$0.09  & 2.38$\pm$0.11  & 7.42$\pm$0.07 
& 3.46$\pm$0.11  & 2.45$\pm$0.08  & 7.19$\pm$0.09  \\
SGMSE+M \cite{sgmse}  
& 3.68$\pm$0.13  & 2.74$\pm$0.10 & 7.57$\pm$0.09    
& 3.45$\pm$0.12  & 2.50$\pm$0.08  & 7.22$\pm$0.11 
& 3.34$\pm$0.10  & 2.38$\pm$0.09  & 7.87$\pm$0.14 \\
\textbf{DisSR} 
&\bfseries 3.75$\pm$0.15 &\bfseries 3.02$\pm$0.09  &\bfseries 7.01$\pm$0.09
&\bfseries 3.52$\pm$0.13 &\bfseries 2.61$\pm$0.12  &\bfseries 6.95$\pm$0.09
&\bfseries 3.57$\pm$0.09 &\bfseries 2.57$\pm$0.11  &\bfseries 6.86$\pm$0.12 \\
\bottomrule
\cmidrule(r){1-10}
\end{tabular}}
\vspace{-0.3cm}
\end{table*}

\subsection{Cross-domain speech restoration}
\para{Diffusion-based speech restoration.} 
We use the speaker-invariant degradation $d_{i}$$\in$$\{d_{i}^{h},d_{i}^{m},d_{i}^{l}\}$ from different middle layer of degradation encoder to condition the reverse process of spectrogram denoiser which is a score-based diffusion model. It is harnessed for noise prediction with the corresponding loss formulated as follows:
\begin{equation}
\begin{aligned}
\mathcal{L}_{\mathrm{SRdiff}}:=\mathbb{E}_{z, \epsilon_{\theta} \sim \mathcal{N}(0,1), d_{i}, t}\left[||\epsilon_{\theta}-\mathcal{S}\left(z_t, t, d_{i}\right)||_2^2\right],
\end{aligned}
\label{diffloss}
\end{equation}
where $\epsilon_{\theta}$ denotes noise to predict, $\mathcal{S}$ represents a conditional UNet-based denoising network, $t\in[0,T]$ and $z_{t}$ signifies time step and speech that is noised at step $t$.


\para{Cross-domain alignment (CDA) training.}
To minimize the degradation gap and boundary offsets between cross-domain speech. We employ CDA training strategy to minimize the distribution shift of different speakers.
Maximum mean discrepancy (MMD) \cite{mmd} is used to determine the equivalence of two distributions, which has been verified to be useful for cross-domain image restoration.
We extend MMD to hierarchical MMD (HMMD), which is applied to all downsampling layers of the degradation encoder to achieve a more suitable shared feature space. HMMD is defined by:

\begin{footnotesize}
\vspace{-0.2cm}
\begin{equation}
    \begin{split}    
    &\mathrm{HMMD}^{2}(d_{i},d_{j})=\\
    &\frac{1}{H\cdot C}\sum_{h=1}^{H}\sum_{c=1}^{C}\parallel\sum_{d_i\in D_i}\phi(d_i^{c})-\sum_{ d_j\in D_j}\phi(d_j^{c})\parallel^{2}	
    \end{split}
    \label{HMMD}
\end{equation}
\vspace{-0.15cm}
\end{footnotesize}

\noindent where $H$ denotes the count of downsampling layers in the degradation encoder, $C$ is the number of degradation types. $\phi(\cdot)$ is the map to reproducing kernel
hilbert space (RKHS). $d_{i}$ and $d_{j}$ are degradation representations extracted from different speakers $D_{i}$ and $D_{j}$ respectively.
We consider the unseen speakers as the cross-domain data, and we divide the cross-domain data into different subdomains based on degradation types and apply HMMD to each subdomain.

\section{Experiment}
\label{exp}
\subsection{Experiment setup}
\textbf{Datasets:} In the pre-training stage, we created simulated low-quality datasets by applying various distortions to LibriTTS. We obtain the pre-trained degradation encoder through degradation representation learning on train-clean-100 set.
In the training stage, we divide train-clean-360 set into training, validation and testing according to the ratio of 8:1:1.
To evaluate our proposed DisSR in the cross-domain scenario, we employ VCTK, AISHELL-3, and JSUT \cite{JSUT} as the cross-domain test set. The sampling rate is set to 22.05kHz. 
\normalsize


\textbf{Training}: 
We consider 6 types of distortions as in \cite{selfremaster,voicefixer}: quantized \& resampled, clipped, band-limited, overdrive, noise, reverb, and create corrupted speech using the open-source pipeline in \cite{selfremaster,voicefixer}, resulting in distorted samples with an SNR in [-5, 20]dB, clipped between [0.1, 0.5], and a bandwidth from 2 kHz to 22.05 kHz. Other operations include: Quantizing speech to 8-bit $\mu$-law and resampling it to 8kHz; Overdriving speech the same as \cite{selfremaster}; Add reverb distortion the same as \cite{voicefixer}. Note that we assume the degradation is the same across speeches and varies between them.


\textbf{Baseline systems}: We compare DisSR with other systems. 1) VoiceFixer \cite{voicefixer}: a unified framework for high-fidelity speech restoration, which can restore speech from multiple distortions. 2) SelfRemaster \cite{selfremaster}: a self-supervised speech restoration method without paired data. 3) SGMSE+M \cite{sgmse}: a diffusion-based generative model which is applied to various speech restoration tasks.

\textbf{Evaluation metrics}: 
We rely on the following metrics, and resample the generated speech if necessary. We adopt include DNSMOS, PESQ-wb, mel-cepstrum distortion (MCD), and structural similarity (SSIM).


\subsection{Hypothesis validation}

\begin{figure}[h]
    \centering
    \vspace{-0.2cm}
    \includegraphics[width=5.8cm]{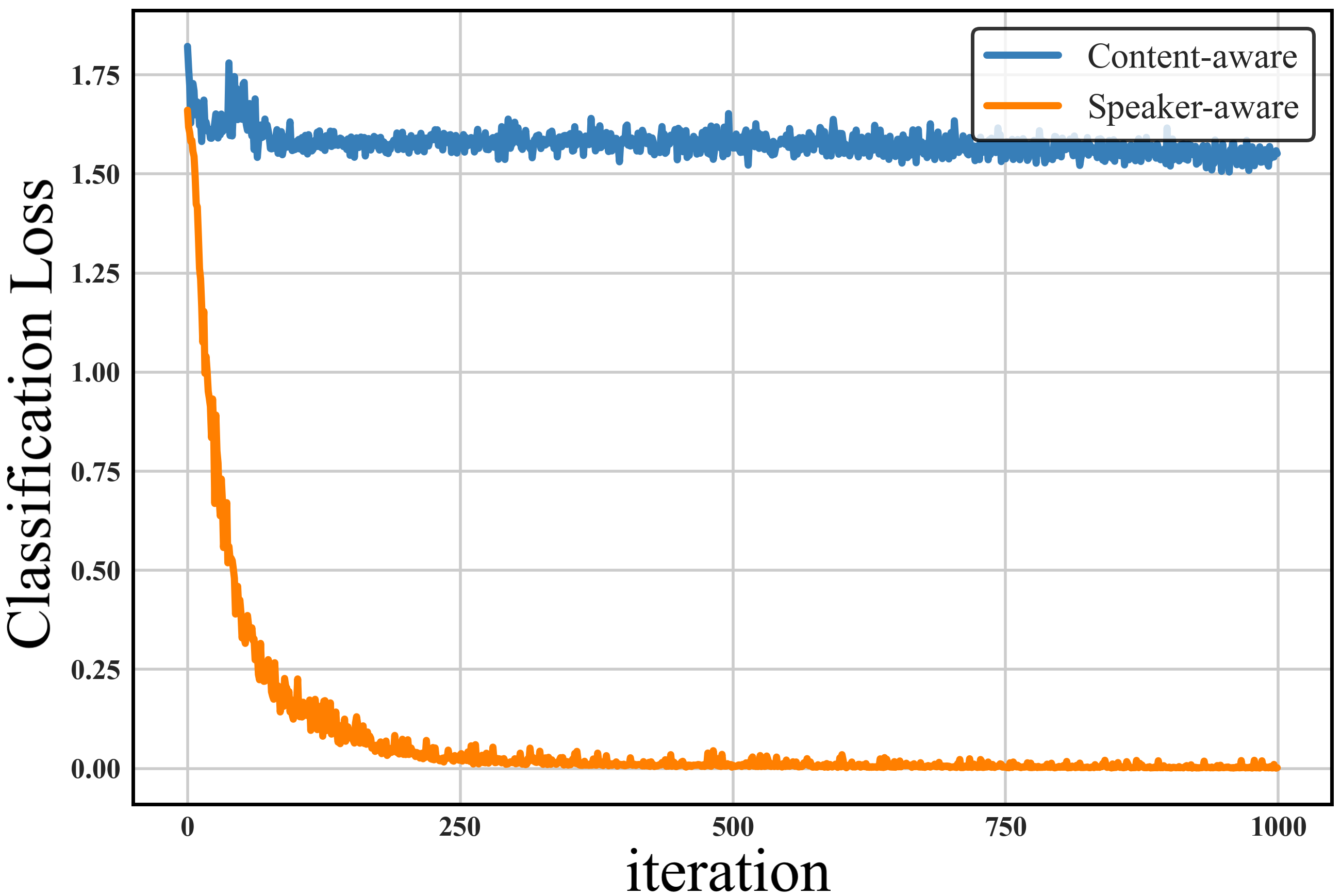}
    \vspace{-0.20cm}
    \caption{Classification loss of degradation classifier.} 
    \vspace{-0.3cm}
    \label{corr_loss}
\end{figure}

We pre-train a degradation-type classifier which adopted ECAPA-TDNN \cite{ecapa-tdnn} to verify our hypotheses. We fed the content and speaker style embedding into the classifier to construct $\mathbf{C}$ontent-aware and $\mathbf{S}$peaker-aware classification loss, and we plot their convergence curves during the training process in \Fref{corr_loss}. 

It can be observed that the convergence speed of Speaker-aware is significantly faster than that of Content-aware, and the loss value reached is lower. It shows that speaker style learns the degradation information more effectively.

\begin{figure}[t]
    \centering
    \includegraphics[width=8.5cm]{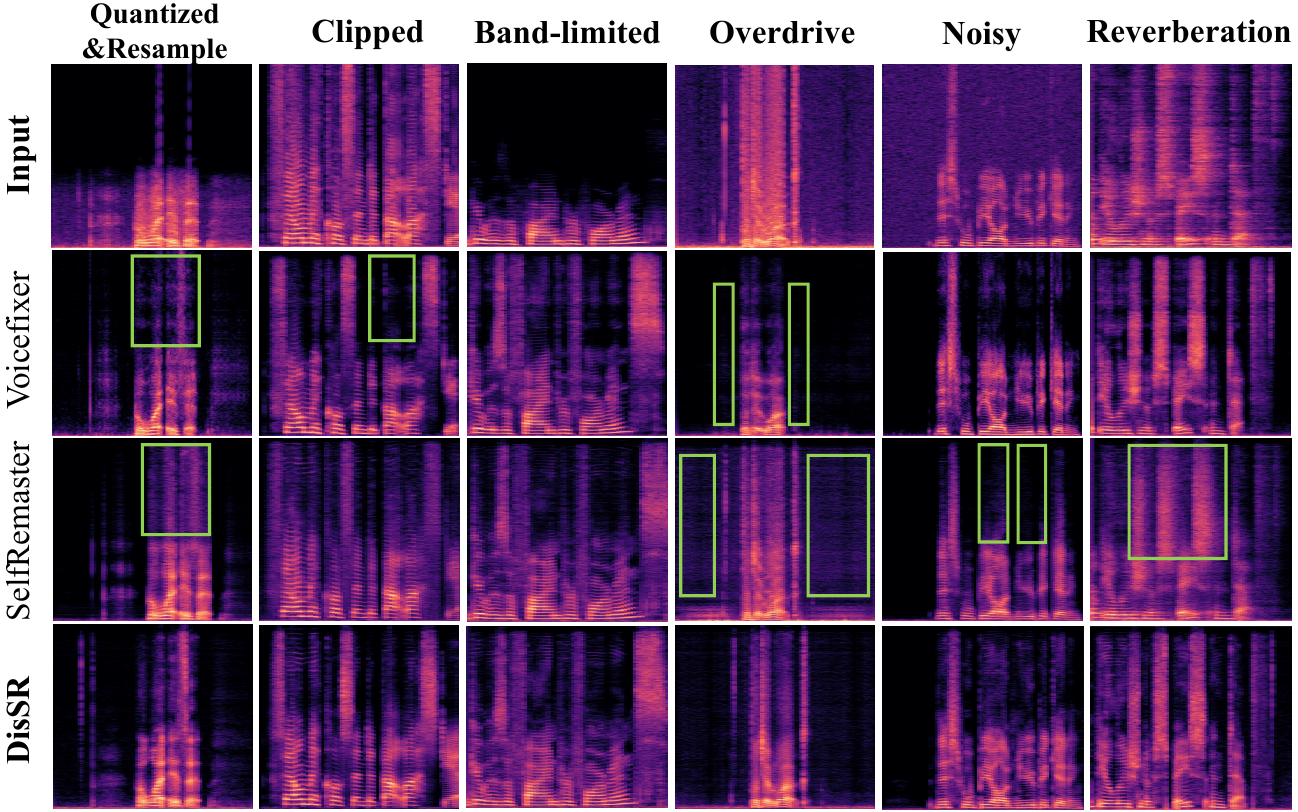}
    \vspace{-0.15cm}
    \caption{Speech restoration results with different methods} 
    \vspace{-0.5cm}
    \label{fig3}
\end{figure}

\subsection{Restoration performance evaluation}

\Tref{audioSR} shows the evaluation results of different methods on various language datasets.
We randomly select 20 sentences from each distortion type and generate restored samples from different models to calculate the DNSMOS, PESQ-wb, and MCD.
Compared with the baseline, DisSR outperforms the baseline model in terms of all metrics. 
It proves that degradation representation served as a conditional cue to guide the diffusion model, which can improve the restoration effect in the face of various speech distortions and the generalization ability for out-of-domain data in real-world scenarios.

\Fref{fig3} supports the findings of \Tref{audioSR} by presenting the spectrograms of distorted input, and outputs of DisSR and VoiceFixer~\cite{voicefixer}, SelfRemaster~\cite{selfremaster}. The figures clearly demonstrate that DisSR performs properly for its designed restoration task, and DisSR has achieved significant results on several tasks.
Quantized \& Resampled is simulated on the basis of the quantization and resampling, and is missing high-frequency bands that are present in the ground-truth high-quality speech, and has distorted low-frequency bands. In DisSR's restored speech, the missing and distorted bands are restored, and it restores high-frequency details better than the baseline. Furthermore, the noise in restored speech illustrates the deficiencies of \cite{selfremaster} in eliminating the overdrive task, and \cite{voicefixer} also has the spectrogram missing problem. In the dereverberation task, DisSR overcomes the high-frequency aliasing problem compared to the baseline.

\subsection{Single restoration task evaluation}
In order to evaluate DisSR's performance on single speech restoration (SSR) task, we compare it with several single-task SSR model as shown in \Tref{tab2}. We utilized composite measurements to analyze the signal distortion (CSIG), background noise (CBAK), and overall quality (COVL)~\cite{TASLP3EVAL}. A higher score on all evaluation metrics signifies improved performance.

In the bandwidth extension task, DisSR achieves the best auditory perceptual quality at 2kHz and 4kHz with minimal background noise and distortion bias, outperforming the single-task model HIFI++.
However, DisSR is still not as good as SGMSE+ in terms of CBAK and StoRM in terms of CSIG in denoising and dereverberation tasks, but it performs well in auditory perceptual quality in terms of COVL. 
These findings highlight DisSR's advantage, attributed to its use of speaker-invariant degradation modeling and CDA training techniques. DisSR improves the domain adaptability of the general restoration model for the SSR task and enables multi-task collaborative work, where multiple restoration tasks benefit from each other rather than conflict with one another.

\begin{table}[t]
  \caption{Comparison with SSR methods on VCTK.}
  \vspace{-0.2cm}
  \label{tab2}
  \centering
  \resizebox{0.47\textwidth}{!}
  {
  \begin{tabular}{@{}ccccc@{}}
    \toprule
    \cmidrule(r){1-5}
     \textbf{Type} & \textbf{Methods} &\textbf{CSIG}$\uparrow$ &\textbf{CBAK}$\uparrow$ &\textbf{COVL}$\uparrow$   \\
    \midrule
    \multirow{4}{*}{Bandwidth extension}
    & HIFI++~\cite{HIFI} (2kHz)     & 3.08$\pm$0.08 & 2.26$\pm$0.06 & 2.87$\pm$0.09    \\
    & DisSR (2kHz)        &\bfseries3.22$\pm$0.06 &\bfseries2.49$\pm$0.09  &\bfseries3.03$\pm$0.07   \\     \cmidrule(l){2-5}  
    & HIFI++~\cite{HIFI} (4kHz)     & 3.51$\pm$0.11 &2.70$\pm$0.09  & 3.14$\pm$0.10   \\
    & DisSR (4kHz)     &\bfseries3.60$\pm$0.09 & \bfseries2.77$\pm$0.08  &\bfseries3.19$\pm$0.11   \\ 
    \midrule
    \multirow{4}{*}{Denoising}
    & DEMUCS~\cite{DEMUCS}      & 3.29$\pm$0.09 &3.58$\pm$0.08  &3.40$\pm$0.07   \\
    & MP-SENet~\cite{lu2023mp}      & 3.41$\pm$0.12 & 3.62$\pm$0.11  &3.47$\pm$0.08  \\
    & SGMSE+~\cite{RichterWLLG23}      & 3.44$\pm$0.05 &\bfseries3.71$\pm$0.10  &3.51$\pm$0.11   \\
    & DisSR              &\bfseries3.48$\pm$0.07 & 3.66$\pm$0.12 &\bfseries 3.58$\pm$0.08    \\ 
    \midrule
    \multirow{3}{*}{Dereverberation}
    & SGMSE+~\cite{RichterWLLG23}  & 3.11$\pm$0.11 & 2.86$\pm$0.11  & 3.07$\pm$0.09   \\
    & StoRM~\cite{lemercier2023storm}  &\bfseries 3.16$\pm$0.12 & 2.91$\pm$0.09  & 3.11$\pm$0.08   \\
    & DisSR       & 3.11$\pm$0.09  &\bfseries2.99$\pm$0.07  &\bfseries 3.15$\pm$0.10   \\ 
    \bottomrule    
  \end{tabular}
  }
  \vspace{-0.3cm}
\end{table}

\subsection{Ablation study}


To evaluate the improvement effect of $\{d_{i}^{h},d_{i}^{m},d_{i}^{l}\}$ on the diffusion-based speech restoration model, we trained DisSR with specific components selectively removed, and the results are presented in \Tref{tab3}.
Comparative metrics are used to evaluate the quality and expressiveness of restored speech.

\begin{table}[h]
  \caption{Ablation study of DisSR on AISHELL-3.}
  \vspace{-0.2cm}
  \label{tab3}
  \centering
  \resizebox{0.45\textwidth}{!}
  {
  \begin{tabular}{ccccc}
    \toprule
    \cmidrule(r){1-5}
     \textbf{Guidance} &\textbf{DNSMOS} $\uparrow$ &\textbf{PESQ-wb} $\uparrow$ &\textbf{MCD} $\uparrow$ &\textbf{SSIM} $\uparrow$    \\
    \midrule
    w/o $d_{i}^{h}$     & /      & -0.05 & -0.13  & -0.03     \\
    w/o $d_{i}^{m}$     & -0.03  & -0.09 & -0.21  & -0.05     \\
    w/o $d_{i}^{l}$     & -0.06  & -0.18 & -0.57  & -0.13     \\
    w/o $\mathcal{L}_{\mathrm{DRL}}$  & -0.14  & -0.22 & -0.33  & -0.11   \\ 
    w/o $\mathcal{L}_{\mathrm{HMMD}}$ & -0.07  & -0.16 & -0.25  & -0.09   \\   
    \bottomrule
  \end{tabular}
  }
  \vspace{-0.2cm}
\end{table}

The decrease in both quality and expressiveness scores shows the significance of speaker-invariant degradation disentanglement and the degradation-prior module. 
Moreover, 
the lack of degradation representation learning and cross-domain training leads to a significant drop in PESQ-wb and MCD, indicating that the domain-invariant degradation-prior guidance plays a crucial role in guiding DisSR to restore speech distortions.

\section{Conclusion}

We proposed DisSR, a multi-task speech restoration model for cross-domain data and multiple distortions. We disentangle low-quality speech into content and speaker style information, and further disentangle distortion features from speaker style information. 
To address the challenges of cross-domain data in real-world scenarios, we designed a degradation-prior module to model speaker-invariant degradation representation through degradation representation learning. Cross-domain alignment training is adopted to degradation disentanglement for improving the distribution shift on cross-domain speech.


\vfill\pagebreak

\bibliographystyle{IEEEbib}
\bibliography{strings,refs}

\end{document}